\documentclass[preprint,aps,prc,nofootinbib]{revtex4}
\usepackage{epsfig}
\usepackage{graphicx}
\usepackage{amsfonts}
\usepackage{amsmath}
\begin{document}

\title{MESON ELECTROMAGNETIC FORM FACTORS}
\date{\today}

\medskip

\author{Stanislav Dubni\v cka}
\address{Institute of Physics, Slovak Academy of Sciences,
Bratislava, Slovak Republic}

\author{Anna Z. Dubni\v ckov\'a}
\address{Department of Theoretical Physics, Comenius University,
Bratislava, Slovak Republic}

\begin{abstract}
     The electromagnetic structure of the pseudoscalar meson nonet
is completely described by the sophisticated $Unitary\&Analytic$
model, respecting all known theoretical properties of the
corresponding form factors.
\end{abstract}

\maketitle

\section{INTRODUCTION}

   All hadrons are compound of constituent quarks. As a consequence in EM
interactions they manifest a non-point-like structure, completely
described by scalar functions $F_i(t)$, called electromagnetic
(EM) form factors (FFs), where $t$ is squared momentum transferred
by the virtual photon $\gamma^*$. If $M \gamma^* \rightarrow M
\Rightarrow F_i(t)$ are called elastic FFs. If $M\gamma^*
\rightarrow A'$ or $\gamma \Rightarrow F_i(t)$ are called
transition FFs.

    According to SU(3) classification there are scalar meson, pseudoscalar meson,
vector meson and tensor meson \cite{nakam} multiplets to be bound
states of light quarks ${u, d, s}$. For a description of their EM
structure we use $Unitary\&Analytic$ ($U\&A$) model \cite{dubdub},
which is a consistent unification of pole and continuum
contributions, depends on effective $t_{in}$ thresholds and the
coupling constant ratios $(f_{MMV}/f_V)$ as free parameters. In
order to determine them numerically one needs a comparison of the
$U\&A$ model with some experimental data. Therefore, farther our
attention is concentrated only to the nonet of pseudoscalar mesons
$\pi^-, \pi^0, \pi^+, K^-, K^0, \bar K^0, K^+, \eta, \eta'$, for
which abundant experimental information exists.

\section{FIRST GENERALLY}

   Since pseudoscalar mesons  $M$  have spin $0^-$
there is only one FF $F_i(t)$ completely describing their EM
structure, which is defined by the parametrization
\begin{equation}
  <p_2|J_\mu(0)|p_1> = e F_M(t) (p_1+p_2)_\mu.
\end{equation}
of the matrix element of the EM current.

   Making use of the transformation $J_\mu(x)$ and also the
one-particle state vectors $<p_2|$ and $|p_1>|$ with regard to all
three discrete $C, P, T$ transformations simultaneously then
$F_M(t)=-F_{\bar M}(t)$ e.g.
 $F_{\pi^+}(t)=-F_{\pi^-}(t); F_{K^+}(t)=-F_{K^-}(t);
 F_{K^0}(t)=-F_{\bar K^0}(t)$.
From the latter it follows for true neutral pseudoscalar mesons
\quad $\pi^0, \eta, \eta'$
  $F_{\pi^0}(t) = F_\eta(t) = F_{\eta'}(t) \equiv 0$
for all values from the interval $-\infty < t < +\infty$.

\section{$U\&A$ MODEL OF MESON EM FFs.}

    There is a general belief that all EM FFs are analytic in t-plane,
besides branch points i.e. cuts on the positive real axis.

    The $U\&A$ model is a consistent unification of
finite number of complex conjugate pairs of poles contributions
and just continua contributions represented by cuts on the
positive real axis.

    Experimental fact of the creation of $\rho, \omega, \phi,
\rho', \omega', \phi', etc.$ in $e^+e^- \to hadrons$ in the first
approximation can be taken into account by the standard $VMD$
model with stable vector mesons
\begin{equation}
   F_M(t) = \sum_V \frac{m^2_V}{m^2_V-t}(f_{MMV}/f_V),
\end{equation}
which automatically respects the asymptotic behavior of
pseudoscalar meson EM FFs
\begin{equation}
   F_M(t)_{|t| \to \infty} \sim t^{-1}
\end{equation}
as predicted by the constituent quark model of hadrons.

    Afterwards the $VMD$ model is unitarized by an
incorporation of two-cut approximation of the analytic properties
of EM FFs with the help of the non-linear transformation
\begin{equation}
   t = t_0 + \frac{4(t_{in}-t_0)}{[1/W(t)-W(t)]^2},
\end{equation}
where $t_0$ is the square-root branch point corresponding to the
lowest possible threshold, $t_{in}$ is an effective square-root
branch point simulating contributions of all higher relevant
thresholds given by the unitarity condition and
\begin{equation}
   W(t)=i\frac{\sqrt{(\frac{t_{in}-t_0}{t_0})^{1/2}+(\frac{t-t_0}{t_0})^{1/2}}-
   \sqrt{(\frac{t_{in}-t_0}{t_0})^{1/2}-(\frac{t-t_0}{t_0})^{1/2}}}
   {\sqrt{(\frac{t_{in}-t_0}{t_0})^{1/2}+(\frac{t-t_0}{t_0})^{1/2}}+
   \sqrt{(\frac{t_{in}-t_0}{t_0})^{1/2}-(\frac{t-t_0}{t_0})^{1/2}}}
\end{equation}
is the conformal mapping of the four-sheeted Riemann surface into
one $W$-plane, to be just inverse to the previous non-linear
transformation.

  As a result every term $\frac{m^2_V}{m^2_V-t}$ in $VMD$
representation is factorized
\begin{eqnarray*}
\frac{m^2_r}{m^2_r-t}=(\frac{1-W^2}{1-W^2_N})^2
\frac{(W_N-W_{r0})(W_N+W_{r0})(W_N-1/W_{r0})(W_N+1/W_{r0})}
{(W-W_{r0})(W+W_{r0})(W-1/W_{r0})(W+1/W_{r0})}
\end{eqnarray*}
into the asymptotic term $(\frac{1-W^2}{1-W^2_N})^2$ completely
determining the asymptotic behavior $\sim t^{-1}$ of EM FF and
into a resonant term
$\frac{(W_N-W_{r0})(W_N+W_{r0})(W_N-1/W_{r0})(W_N+1/W_{r0})}
  {(W-W_{r0})(W+W_{r0})(W-1/W_{r0})(W+1/W_{r0})}$,
for $\mid t \mid\to \infty$ turning out to real constant. The
subindex $"0"$ means that still stable vector-mesons are
considered.

   Generally one can prove if $m^2_r-\Gamma^2_r/4 < t_{in} \Rightarrow W_{r0}=-W^*_{r0}$
and if $m^2_r-\Gamma^2_r/4 > t_{in} \Rightarrow
W_{r0}=1/W^*_{r0}$, which lead in the first case to the expression
\begin{eqnarray*}
\frac{m^2_r}{m^2_r-t}=\left(\frac{1-W^2}{1-W^2_N}\right )^2
\frac{(W_N-W_{r0})(W_N-W^*_{r0})(W_N-1/W_{r0})(W_N-1/W^*_{r0})}
{(W-W_{r0})(W-W^*_{r0})(W-1/W_{r0})(W-1/W^*_{r0})}
\end{eqnarray*}
and in the second case to the following expression
\begin{eqnarray*}
\frac{m^2_r}{m^2_r-t}=\left (\frac{1-W^2}{1-W^2_N}\right )^2
\frac{(W_N-W_{r0})(W_N-W^*_{r0})(W_N+W_{r0})(W_N+W^*_{r0})}
{(W-W_{r0})(W-W^*_{r0})(W+W_{r0})(W+W^*_{r0})}.
\end{eqnarray*}

   Finally, introducing the non-zero widths of resonances by a formal substitution
$m^2_r \rightarrow (m_r - \Gamma_r/2)^2$ i.e. simply one has to
rid of $0$ in subindices, one gets, when the resonance is below
$t_{in}$
\begin{eqnarray*}
\frac{m^2_r}{m^2_r-t} \rightarrow \left
(\frac{1-W^2}{1-W^2_N}\right
)^2\frac{(W_N-W_{r})(W_N-W^*_{r})(W_N-1/W_{r})(W_N-1/W^*_{r})}
{(W-W_{r})(W-W^*_{r})(W-1/W_{r})(W-1/W^*_{r})}
\end{eqnarray*}
and when the resonance is beyond $t_{in}$
\begin{eqnarray*}
\frac{m^2_r}{m^2_r-t} \rightarrow \left
(\frac{1-W^2}{1-W^2_N}\right )^2
\frac{(W_N-W_{r})(W_N-W^*_{r})(W_N+W_{r})(W_N+W^*_{r})}
  {(W-W_{r})(W-W^*_{r})(W+W_{r})(W+W^*_{r})}
\end{eqnarray*}
where no more equality can be used in these relations.

\begin{figure}
\includegraphics[scale=0.4]{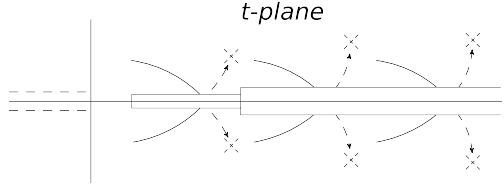}
\caption{Analytic properties of charged pion EM FF.}
\label{Fig.1}
\end{figure}
    Consequently, the $U\&A$ model of meson EM structure takes the
form
\begin{eqnarray*}
    F_P[W(t)]=\left (\frac{1-W^2}{1-W^2_N}\right )^2 \times\\
 \times \Bigg \{ \sum_i \frac{(W_N-W_{i})(W_N-W^*_{i})(W_N-1/W_{i})(W_N-1/W^*_{i})}
  {(W-W_{i})(W-W^*_{i})(W-1/W_{i})(W-1/W^*_{i})}(f_{iPP}/f_i)+\\
 + \sum_j \frac{(W_N-W_{j})(W_N-W^*_{j})(W_N+W_{j})(W_N+W^*_{j})}
  {(W-W_{j})(W-W^*_{j})(W+W_{j})(W+W^*_{j})}(f_{jPP}/f_j)\Bigg \}
\end{eqnarray*}

which is analytic in the whole complex Fig.~1 besides two cuts
on the positive real axis.

\section{NOW ONE BY ONE}

\underline{$\pi^{\pm}$}:  The analytic properties of $F_\pi(t)$
are in Fig.~2. In comparison with expression $F_P[W(t)]$ there
is additional left-hand cut on the II.Riemann sheet.

The latter is explained by the following way. Starting from the
elastic unitarity condition for $F_\pi(t)$
$\frac{1}{2i}\{F_\pi(t+i\varepsilon) - F^*_\pi(t+i\varepsilon)\} =
   A^{1*}_1(t+i\varepsilon).F_\pi(t+i\varepsilon)$
one can derive the expression for pion EM FF on the II.Riemann
sheet $[F_\pi(t)]^{II.}=\frac{F_\pi(t)}{1+2iA^1_1(t)}$ where
$A^1_1(t)$ is the  $P$-wave isovector $\pi\pi$-scattering
amplitude, the analytic properties of which consist of right-hand
unitary cut $4m^2_\pi < t < \infty$ and of left-hand dynamical cut
$-\infty < t < 0.$. Taking into account the fact that the
contribution of any cut in Pad'e approximation can be represented
by alternating zeros and poles on the place of the cut then we do
it in $U\&A$ model of $F_\pi[W(t)]$.

   From the same elastic unitarity condition and
$\delta^1_1(t)_{q \to 0}\sim a^1_1q^3$ one gets the threshold
behavior of $Im F_\pi(t)$ to be transformed into 3 threshold
conditions
   $Im F_\pi(t)_{q=0}=\frac{dIm F_\pi(t)}{dq}_{q=0}=\frac{d^2Im
   F_\pi(t)}{dq}_{q=0}\equiv 0$,
which reduce a number of $(f_{v\pi\pi}/f_v)$ as free parameters.

   Taking into account both these notes and also the
normalization explicitly one gets the $U\&A$ pion EM FF model
\cite{dub}
\begin{eqnarray*}
F_\pi[W(t)]=\left (\frac{1-W^2}{1-W^2_N}\right )^2 \frac{(W-W_z)(W_N-W_p)}{(W_N-W_z)(W-W_p)}\times\\
 \times \Bigg \{\frac{(W_N-W_\rho)(W_N-W^*_\rho)(W_N-1/W_\rho)(W_N-1/W^*_\rho)}
  {(W-W_\rho)(W-W^*_\rho)(W-1/W_\rho)(W-1/W^*_\rho)}(f_{\rho\pi\pi}/f_\rho)+\\
 + \sum \limits_{v=\rho',\rho''} \frac{(W_N-W_v)(W_N-W^*_v)(W_N+W_v)(W_N+W^*_v)}
  {(W-W_v)(W-W^*_v)(W+W_v)(W+W^*_v)}(f_{v\pi\pi}/f_v)\Bigg \}
\end{eqnarray*}
with $(f_{\rho'\pi\pi}/f_{\rho'})=\frac{\frac{N_{\rho''}}{\mid
W_{\rho''}\mid^4}}{\frac{N_{\rho'}}{\mid W_{\rho'}\mid^4}
-\frac{N_{\rho''}}{\mid
W_{\rho''}\mid^4}}-\frac{\frac{N_{\rho''}}{\mid
W_{\rho''}\mid^4}+(1+2\frac{W_z.W_p}{W_z-W_p}.Re [W_\rho(1+\mid
W_\rho\mid^{-2})])N_\rho}{\frac{N_{\rho'}}{\mid W_{\rho'}\mid^4}
-\frac{N_{\rho''}}{\mid
W_{\rho''}\mid^4}}(f_{\rho\pi\pi}/f_\rho)$\\
$(f_{\rho''\pi\pi}/f_{\rho''})=1-\frac{\frac{N_{\rho''}}{\mid
W_{\rho''}\mid^4}}{\frac{N_{\rho'}}{\mid W_{\rho'}\mid^4}
-\frac{N_{\rho''}}{\mid W_{\rho''}\mid^4}}+
[\frac{\frac{N_{\rho''}}{\mid
W_{\rho''}\mid^4}+(1+2\frac{W_z.W_p}{W_z-W_p}.Re [W_\rho(1+\mid
W_\rho\mid^{-2})])N_\rho}{\frac{N_{\rho'}}{\mid W_{\rho'}\mid^4}
-\frac{N_{\rho''}}{\mid
W_{\rho''}\mid^4}}-1](f_{\rho\pi\pi}/f_\rho)$

   Due to the $\rho - \omega$ interference effect one has to
carry out the fit of existing data by
   $\mid F_\pi[W(t)]+ R.e^{i\phi}\frac{m^2_\omega}{m^2_\omega -t-im_\omega
   \Gamma_\omega}\mid$
with
   $\phi=arc tg \frac{m_\rho\Gamma_\rho}{m^2_\rho-m^2_\omega}$.

   A description of existing data in space-like and time-like
regions simultaneously with parameters values $t_{in}=(1.296 \pm
0.011) GeV^2$; $R=0.0123 \pm 0.0032$; $W_z=0.3722 \pm 0.0008$;
$W_p=0.5518 \pm 0.0003$; $m_\rho=(759.26 \pm 0.04) MeV$;
$\Gamma_\rho=(141.90\pm0.13)MeV$; $m_{\rho'}=(1395.9 \pm
54.3)MeV$; $\Gamma_{\rho'}=(490.9 \pm 118.8) MeV$
$m_{\rho''}=(1711.5 \pm 63.6) MeV$; $\Gamma_{\rho''}=(369.5 \pm
112.7) MeV$; $(f_{\rho\pi\pi}/f_\rho)=1.0063 \pm 0.0024$;
$\chi^2/ndf=1.58$; is presented in Fig.~2.

\begin{figure}
\includegraphics[scale=0.4]{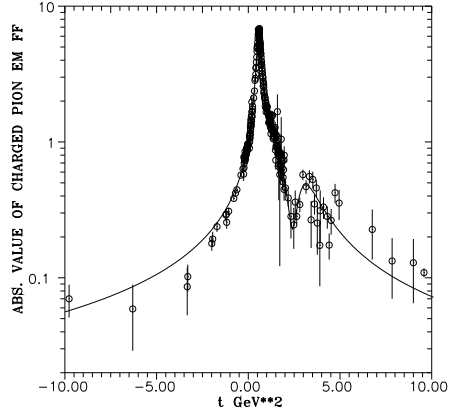}
\caption{Prediction of pion EM FF behavior by $U\&A$ model.}
\label{Fig.2}
\end{figure}

\underline{$K^\pm, K^0$}:

     The $K^+$ and $K^0$ belong to the same isomultiplet with
$I=1/2$. Then one can introduce, generally, the EM current of $K$,
which splits into sum of isotopic scalar and isotopic vector.

   The corresponding FFs suitable for a construction of the $U\&A$
models are
  $F^s_K(t)=\frac{1}{2}[F_{K^+}(t)+F_{K^0}(t)]$ \quad
  $F_{K^+}(t)=F^s_K(t)+F^v_K(t)$\\
  $F^v_K(t)=\frac{1}{2}[F_{K^+}(t)-F_{K^0}(t)]$ \quad
  $F_{K^0}(t)=F^s_K(t)-F^v_K(t)$\\
from where the normalizations
  $F^s_K(0)=F^v_K(0)=\frac{1}{2}$; \quad
  $F_{K^+}(0)=1$; \quad $F_{K^0}(0)=0$;
follow. The specific 6 resonance ($\rho, \omega, \phi, \rho',
\phi', \rho''$) $U\&A$ model of the kaon EM structure has the form
\cite{ddl}
\begin{eqnarray}
\nonumber
  F^s_K[V(t)]=\Bigg(\frac{1-V^2}{1-V^2_N}\Bigg)^2
  \Bigg[\frac{1}{2}\frac{(V_N-V_\omega)(V_N-V^*_\omega)(V_N-1/V_\omega)(V_N-1/V^*_\omega)}
  {(V-V_\omega)(V-V^*_\omega)(V-1/V_\omega)(V-1/V^*_\omega)}+\\\nonumber
 +\Bigg\{\frac{(V_N-V_\phi)(V_N-V^*_\phi)(V_N-1/V_\phi)(V_N-1/V^*_\phi)}
  {(V-V_\phi)(V-V^*_\phi)(V-1/V_\phi)(V-1/V^*_\phi)}-\\
  -\frac{(V_N-V_\omega)(V_N-V^*_\omega)(V_N-1/V_\omega)(V_N-1/V^*_\omega)}
  {(V-V_\omega)(V-V^*_\omega)(V-1/V_\omega)(V-1/V^*_\omega)}\Bigg\}(f_{\phi
  KK}/f_\phi)+\\\nonumber
 + \Bigg\{\frac{(V_N-V_{\phi'})(V_N-V^*_{\phi'})(V_N-1/V_{\phi'})(V_N-1/V^*_{\phi'})}
  {(V-V_{\phi'})(V-V^*_{\phi'})(V-1/V_{\phi'})(V-1/V^*_{\phi'})}-\\\nonumber
  -\frac{(V_N-V_\omega)(V_N-V^*_\omega)(V_N-1/V_\omega)(V_N-1/V^*_\omega)}
  {(V-V_\omega)(V-V^*_\omega)(V-1/V_\omega)(V-1/V^*_\omega)}\Bigg\}(f_{\phi'
  KK}/f_{\phi'})\Bigg]\nonumber
\end{eqnarray}

\begin{eqnarray}
 \nonumber
  F^v_K[W(t)]=\Bigg(\frac{1-W^2}{1-W^2_N}\Bigg)^2
  \Bigg[\frac{1}{2}\frac{(W_N-W_\rho)(W_N-W^*_\rho)(W_N-1/W_\rho)(W_N-1/W^*_\rho)}
  {(W-W_\rho)(W-W^*_\rho)(W-1/W_\rho)(W-1/W^*_\rho)}+\\\nonumber
 +\Bigg\{\frac{(W_N-W_{\rho'})(W_N-W^*_{\rho'})(W_N-1/W_{\rho'})(W_N-1/W^*_{\rho'})}
  {(W-W_{\rho'})(W-W^*_{\rho'})(W-1/W_{\rho'})(W-1/W^*_{\rho'})}-\\
  -\frac{(W_N-W_\rho)(W_N-W^*_\rho)(W_N-1/W_\rho)(W_N-1/W^*_\rho)}
  {(W-W_\rho)(W-W^*_\rho)(W-1/W_\rho)(W-1/W^*_\rho)}\Bigg\}(f_{\rho'
  KK}/f_{\rho'})+\\\nonumber
 + \Bigg\{\frac{(W_N-W_{\rho''})(W_N-W^*_{\rho''})(W_N-1/W_{\rho''})(W_N-1/W^*_{\rho''})}
  {(W-W_{\rho''})(W-W^*_{\rho''})(W-1/W_{\rho''})(W-1/W^*_{\rho''})}-\\\nonumber
  -\frac{(W_N-W_\rho)(W_N-W^*_\rho)(W_N-1/W_\rho)(W_N-1/W^*_\rho)}
  {(W-W_\rho)(W-W^*_\rho)(W-1/W_\rho)(W-1/W^*_\rho)}\Bigg\}(f_{\rho''
  KK}/f_{\rho''})\Bigg]\nonumber
\end{eqnarray}
   Both functions are analytic in the whole complex
$t$-planes besides two cuts on the positive real axis, generated
by $t^s_0=9 m^2_\pi$ and $t^s_{in}$ in $F^s_K[V(t)]$ and by
$t^v_0=4 m^2_\pi$ and $t_{in}$ in $F^v_K[W(t)]$. They are  real on
the whole real negative axis up to positive values $t^s_0=9
m^2_\pi$ and $t^v_0=4 m^2_\pi$, respectively, automatically
normalized to $1/2$ with $Im F^s_K(t)\neq 0$ and $Im F^v_K(t)\neq
0$, starting from $9 m^2_\pi$ and $4 m^2_\pi$, respectively, as it
is required by the unitarity conditions. They possess complex
conjugate pairs of poles on unphysical sheets of the Riemann
surface, corresponding to considered vector-mesons with quantum
numbers of the photon.

  A simultaneous reproduction of all existing kaon EM FF
data by the $U\&A$ models is presented in Fig.~3 and Fig.~4

\begin{figure}
\includegraphics[scale=0.35]{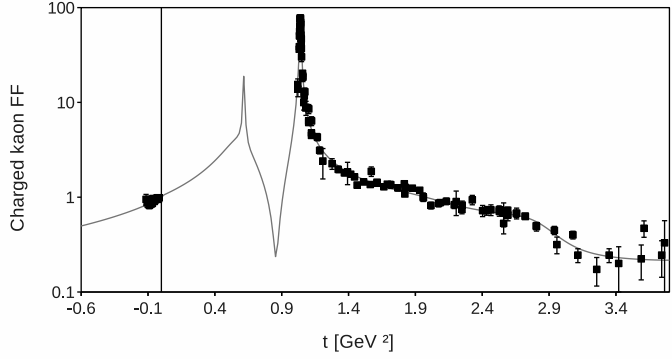}
\caption{Prediction of charge kaon EM FF behavior by $U\&A$
model.}
\label{Fig.3}
\end{figure}
\begin{figure}
\includegraphics[scale=0.35]{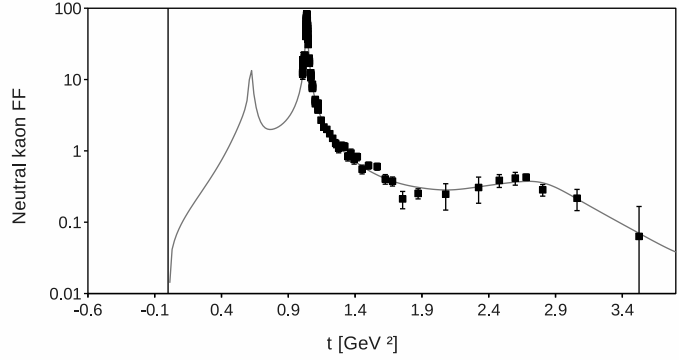}
\caption{Prediction of neutral kaon EM FF behavior by $U\&A$
model.}
\label{Fig.4}
\end{figure}
and the following values of free parameters of the model have been
determined - $m_\rho, \Gamma_\rho, m_\omega, \Gamma_\omega$ are
fixed at the TABLE values. $q^s_{in}=\sqrt{(t^s_{in}-9)/9}=2.2326
[m_\pi]$; $q^v_{in}=\sqrt{(t^v_{in}-4)/4}=6.6721[m_\pi]$;
$(f_{\omega KK}/f_\omega)=0.14194$;
$(f_{\rho KK}/f_\rho)=0.5615$;\\
$m_\phi=7.2815[m_\pi]$; $m_{\rho'}=10.3940[m_\pi]$;
$\Gamma_\phi=0.03733[m_\pi]$;
$\Gamma_{\rho'}=1.6284[m_\pi]$;\\
$(f_{\phi KK}/f_\phi)=0.4002$;
$(f_{\rho'KK}/f_{\rho'})=-.3262$;\\
$m_{\phi'}=11.8700[m_\pi]$; $m_{\rho''}=13.5650[m_\pi]$;
$\Gamma_{\phi'}=1.3834[m_\pi]$;
$\Gamma_{\rho''}=3.3313[m_\pi]$;\\
$(f_{\phi'KK}/f_{\phi'})=-.04214$;
$(f_{\rho''KK}/f_{\rho''})=-.02888$

\underline{What about $\pi^0, \eta, \eta'$}:

   They are true neutral particles and then their elastic EM FFs
$F_{\pi^0}(t)=0$; $F_\eta(t)=0$; $F_{\eta'}(t)=0$ i.e. these
particles are point-like according to EM interactions.

   However, one can define nonzero single FF for each $\gamma^*
\to \gamma P$ transition by a parametrization of the matrix
element of the EM current
   $<P(p)\mid J^{EM}_\mu\mid
   0>=\varepsilon_{\mu\nu\alpha\beta}p^\nu\epsilon^\alpha k^\beta
   F_{\gamma P}(q^2)$ with $\epsilon^\alpha$ to be the polarization vector of
   $\gamma$, and $\varepsilon_{\mu\nu\alpha\beta}$ is antisymmetric
   tensor.

   The transition FFs are related to corresponding cross
sections\\
$\sigma_{tot}(e^+e^- \to P \gamma)=\frac{\pi
  \alpha^2}{6}(1-\frac{m^2_P}{t})^3 \mid F_{P\gamma}(t)\mid^2$\\
giving experimental data on $F_{\pi^0\gamma}(t),
F_{\eta\gamma}(t)$ and $F_{\eta'\gamma}(t)$ in $t>0$ region.

   A straightforward calculation of $F_{P\gamma}(t)$ in $QCD$
is impossible. One has to construct sophisticated phenomenological
models.

   In a construction of the $U\&A$ model it is again suitable to
split $F_{P\gamma}(t)$ into two terms depending on the isotopic
character of the photon
   $F_{P\gamma}(t)=F^{I=0}_{P\gamma}(t)+F^{I=1}_{P\gamma}(t)$
where $F^{I=0}_{P\gamma}(t)$ is saturated by isoscalar
vector-mesons $\omega, \phi, \omega',\phi'$ etc. and
$F^{I=1}_{P\gamma}(t)$ is saturated by  isovector vector-mesons
$\rho, \rho', \rho''$ etc. However, there is a question how many
vector-meson resonances have to be taken into account. It is
prescribed by the existing data interval on the corresponding FF
in $t>0$ region.

   The data on $\pi^0\gamma$ transition FF allow to consider all
$3$ ground state vector mesons, $\rho(770), \omega(782),
\phi(1020)$ and also $\omega'(1420)$ and $\rho'(1450)$, in order
to construct automatically normalized $U\&A$ models.

   With the aim of obtaining comparable results, the same
number of resonances is considered also for $\eta$ and $\eta'$.

   In the analysis the resonance parameters are fixed at the TABLE values,
the normalization of FFs are $F_{P\gamma}(0)=\frac{2}{\alpha
m_P}\sqrt{\frac{\Gamma(P \to \gamma\gamma)}{\pi m_P}}$ where
$\Gamma(P \to \gamma\gamma)$ are fixed at the world averaged
values from TABLE.

   The $F_{P\gamma}(t)$  FFs are analytic
in $t$ - plane besides the cut from $t=m^2_{\pi^0}$ up to
$+\infty$. Then the $U\&A$ model of $F_{P\gamma}(t)$ takes the
form \cite{ddl2}\\
$F^{I=0}_{P\gamma}[V(t)]=\Bigg(\frac{1-V^2}{1-V^2_N}\Bigg)^2
{\frac{1}{2}F_{P\gamma}(0)H(\omega')+
[L(\omega)-H(\omega')]a_\omega+[H(\phi)-H(\omega')]a_\phi\}}$\\
$F^{I=1}_{P\gamma}[W(t)]=\Bigg(\frac{1-W^2}{1-W^2_N}\Bigg)^2
\{\frac{1}{2}F_{P\gamma}(0)H(\rho')+[L(\rho)-H(\rho')]a_\rho\}$\\
with
$L(\omega)=\frac{(V_N-V_\omega)(V_N-V^*_\omega)(V_N-1/V_\omega)(V_N-1/V^*_\omega)}
{(V-V_\omega)(V-V^*_\omega)(V-1/V_\omega)(V-1/V^*_\omega)}$;
$H(i)=\frac{(V_N-V_i)(V_N-V^*_i)(V_N+V_i)(V_N+V^*_i)}
{(V-V_i)(V-V^*_i)(V+V_i)(V+V^*_i)}$; $i=\phi, \omega'$

$L(\rho)=\frac{(W_N-W_\rho)(W_N-W^*_\rho)(W_N-1/W_\rho)(W_N-1/W^*_\rho)}
{(W-W_\rho)(W-W^*_\rho)(W-1/W_\rho)(W-1/W^*_\rho)}$;
$H(\rho')=\frac{(W_N-W_{\rho'})(W_N-W^*_{\rho'})(W_N+W_{\rho'})(W_N+W^*_{\rho'})}
{(W-W_{\rho'})(W-W^*_{\rho'})(W+W_{\rho'})(W+W^*_{\rho'})}$

and the normalization points $V(0)=V_N,$ \quad $W(0)=W_N$.

  The model depends on $5$ free parameters $t^s_{in}, t^v_{in}, a_j=(f_{\gamma P j}/f_j),$ \quad $j=\rho,
   \omega, \phi$ determined in an optimal description of existing data.

\underline{for $\pi^0$:} see Fig.~5

$q^s_{in}=5.5210 \pm 0.0084$; $q^v_{in}=5.61220 \pm 0.1414$; $a_\omega=0.0063 \pm 0.0013$;\\
$a_\rho=0.0212 \pm 0.0006$; $a_\phi=-.0004 \pm 0.0001$;
$\chi^2/ndf=121/75=1.61$

\begin{figure}
\includegraphics[scale=0.35]{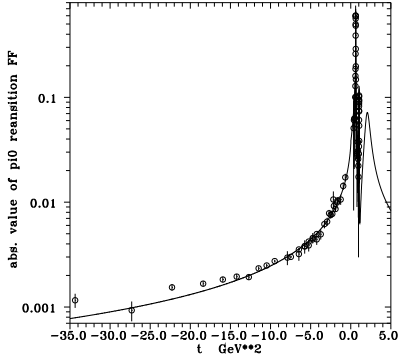}
\caption{Prediction of $\pi^0\gamma$ transition EM FF behavior by
$U\&A$ model.}
\label{Fig.5}
\end{figure}
\begin{figure}
\includegraphics[scale=0.35]{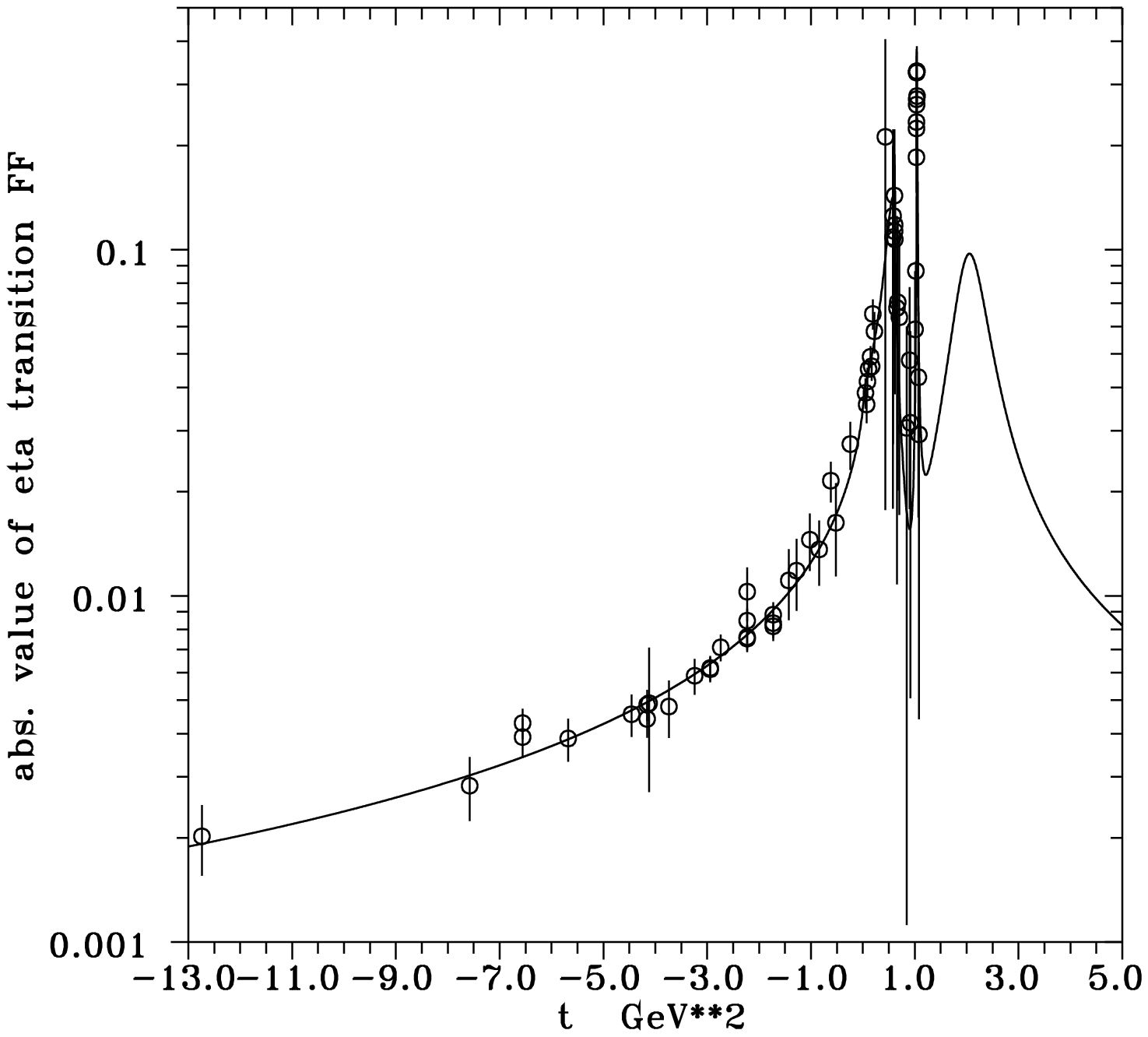}
\caption{Prediction of $\eta\gamma$ transition EM FF behavior by
$U\&A$ model.}
\label{Fig.6}
\end{figure}
\begin{figure}
\includegraphics[scale=0.35]{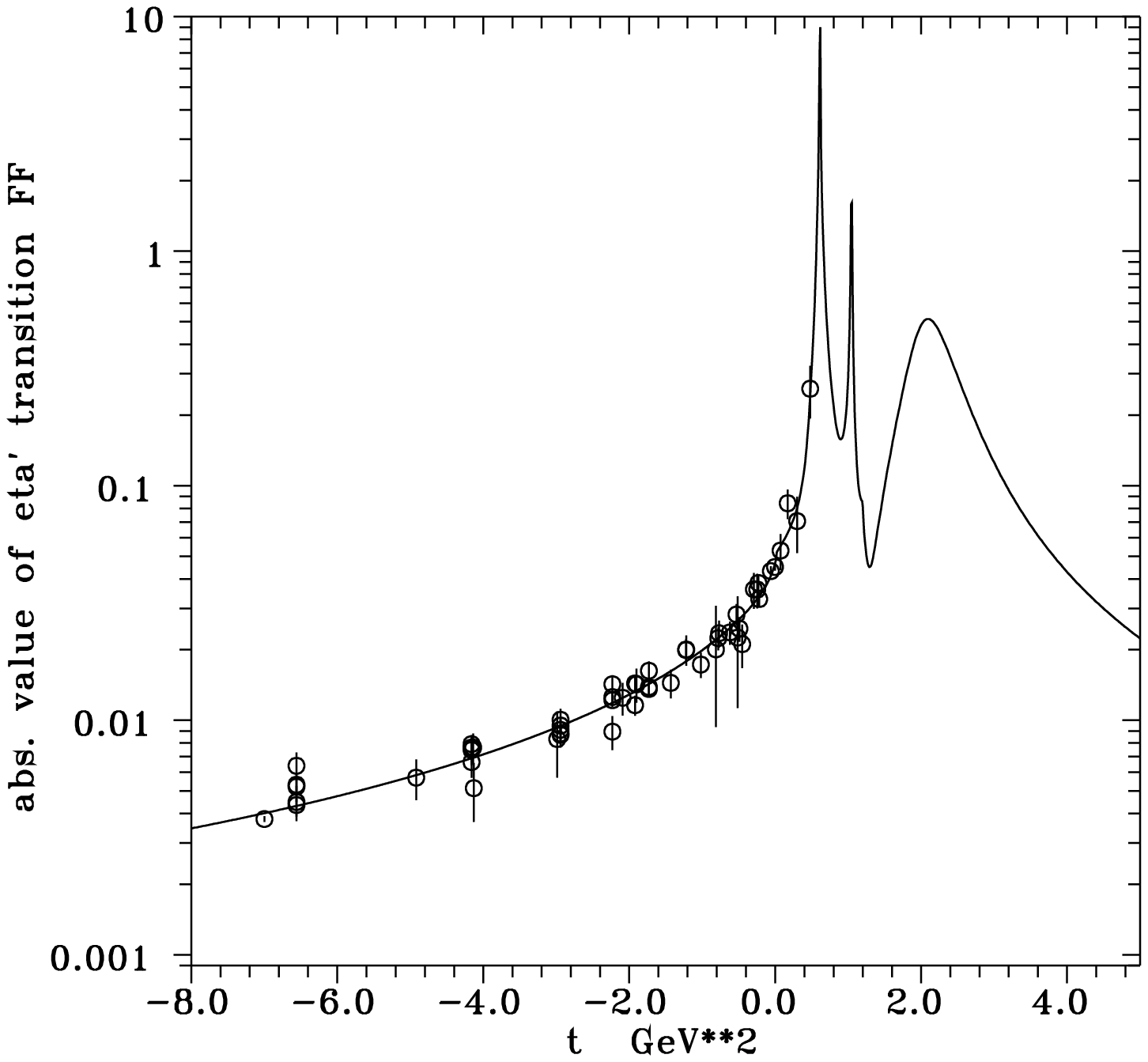}
\caption{Prediction of $\eta'\gamma$ transition EM FF behavior by
$U\&A$ model.}
\label{Fig.7}
\end{figure}

\underline{for $\eta$:} see Fig.~6

$q^s_{in}=6.7104 \pm 0.0190$; $q^v_{in}=5.5006 \pm 0.0632$;
$a_\omega=0.0002 \pm 0.0014$;\\ $a_\rho=0.0250 \pm 0.0013$;
$a_\phi=-.0020 \pm 0.0003$;  $\chi^2/ndf=52/52=1.00$

\underline{for $\eta'$:} see Fig.~7

$q^s_{in}=5.5366 \pm 0.0891$; $q^v_{in}=7.7554 \pm 0.0158$;
$a_\omega=-.1134 \pm 0.0078$;\\ $a_\rho=0.1241 \pm 0.0026$;
$a_\phi=0.0098 \pm 0.0091$; $\chi^2/ndf=59/50=1.18$

\section{CONCLUSIONS}

  We have investigated EM structure of pseudoscalar mesons to be
described by the corresponding EM FFs. Since there is no
possibility to describe the latter in the framework of $QCD$, the
universal $U\&A$ models have been elaborated.

  More or less successful description of all existing data on the whole complete nonet
of pseudoscalar mesons $\pi^-, \pi^0, \pi^+, K^-, K^0, \bar K^0,
K^+, \eta, \eta'$ has been achieved in space-like and time-like
regions simultaneously.

The work was supported in part by Slovak Grant Agency for Sciences VEGA, Grant 2/0009/10.

\end{document}